\documentclass[journal=jacsat,manuscript=article]{achemso}

\usepackage[version=3]{mhchem} 
\usepackage{amsmath}

\author{R.J. Wijnhorst}
\affiliation[University of Amsterdam]
{Van der Waals-Zeeman Institute, Institute of Physics, University of Amsterdam, The Netherlands}
\author{M. Prat}
\affiliation[Université de Toulouse]
{Institut de Mécanique des Fluides de Toulouse (IMFT), Université de Toulouse, CNRS, Toulouse, France }
\author{N. Shahidzadeh}
\email{n.shahidzadeh@uva.nl}

\title[An \textsf{achemso} demo]
  {Self-similarity in creeping salt crystallization }


\begin{document}

\begin{tocentry}
\includegraphics[width=9cm]{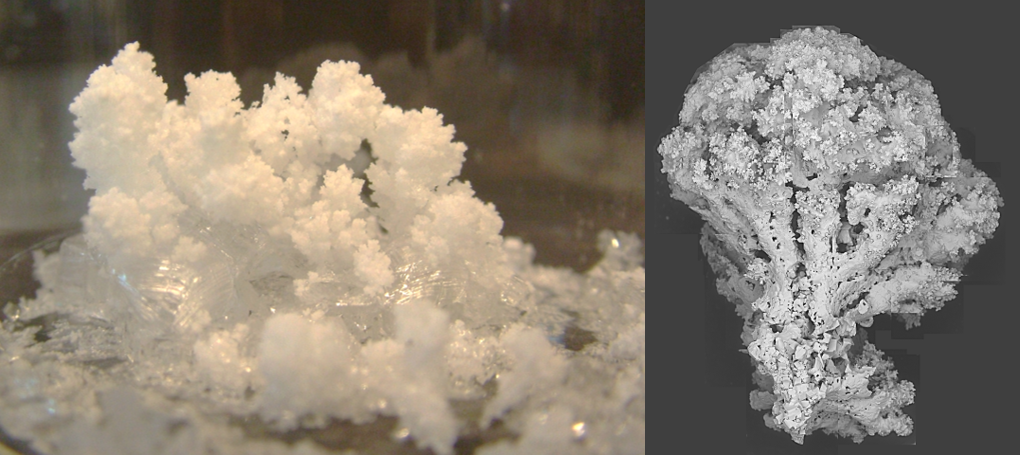}





\end{tocentry}

\begin{abstract}

The self-amplifying creeping of salts can produce striking macroscopic structures, such as desert roses in arid regions and salt pillars near saline lakes. While these formations are visually remarkable, salt crystallization—often seen as efflorescence on surfaces—also poses significant challenges for cultural heritage conservation, materials science, and soil management. In this study, we investigate the mechanisms underlying self-organized crystallization within efflorescence deposits.
Our findings reveal that these porous salt deposits exhibit pronounced self-similarity, with the crystallization process recurring at multiple length scales. This results in smaller replicas of the overall structure nested within larger ones, creating fractal geometries similar to those found in cauliflower and broccoli. By performing controlled evaporation experiments and microscale analysis using advanced imaging techniques combined with fractal dimension analysis, we uncover the hierarchical and size-controlled precipitation of cubic microcrystals within the porous efflorescence. Furthermore, we develop a hierarchical growth model demonstrating that the ultimate height of the macroscopic salt deposit is primarily determined by the initial mass of salt, rather than by the interplay of capillary and viscous forces when salt solution flows within the porous salt structure.
These insights provide a deeper understanding of mineral precipitation and growth, illustrating how natural self-organization at the microscale can give rise to spectacular hierarchical structures. Our results have important implications for managing salt-related phenomena in both natural and engineered environments, and for addressing practical challenges in conservation and materials science.

\end{abstract}

\begin{figure}
\centering
\includegraphics[width=1\textwidth]{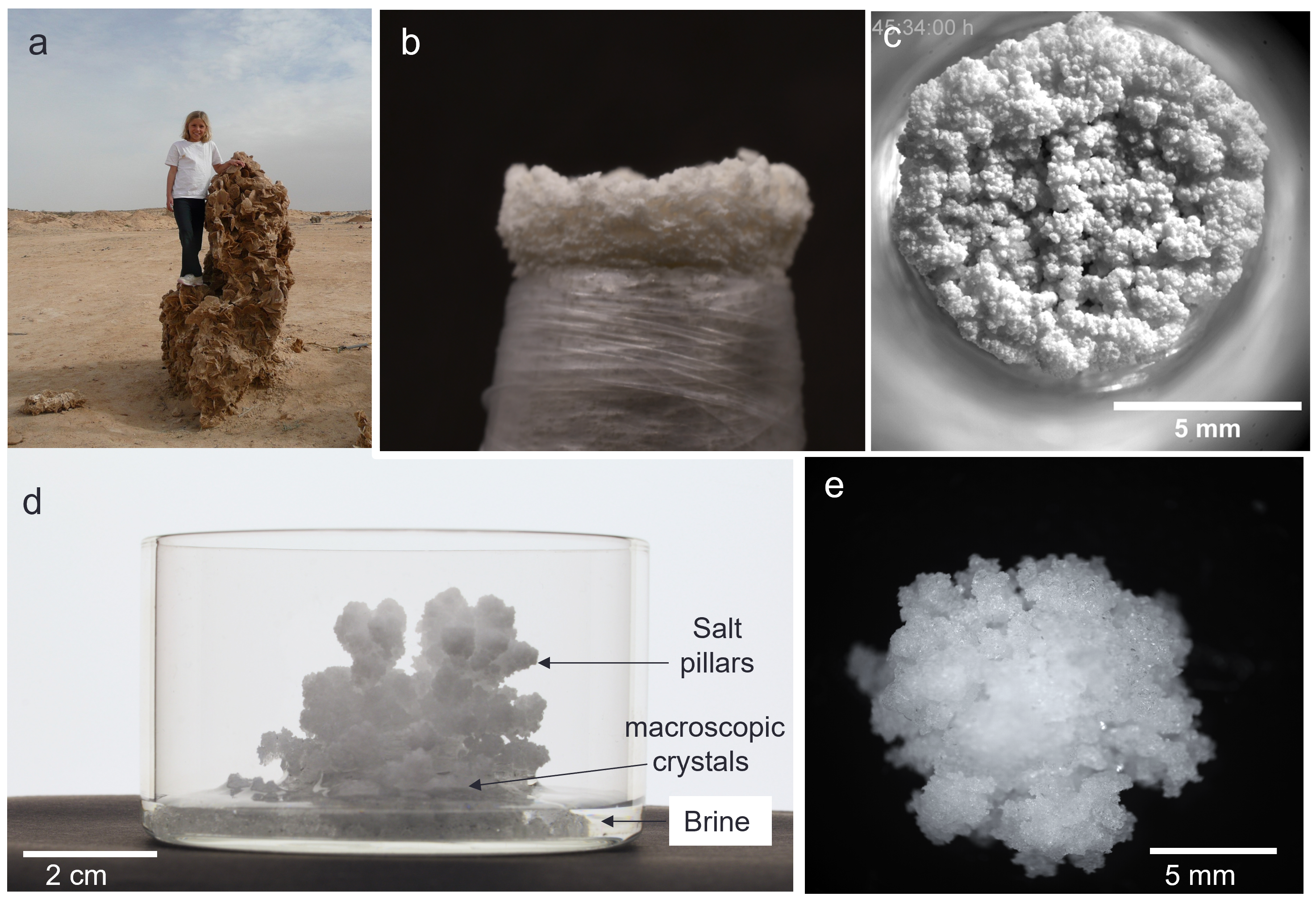}
  \caption{a) Desert rose in Tunisia (courtesy of Romane Le Dizes Castell), b and c) salt efflorescence grown on top of porous sandstone seen from the side and the top, c) Macroscopic NaCl Pillars grown from evaporating brine solution d) top view of one of the branches of the pillar taken off showing a cauliflower shaped of the polycristalline structure}
  \label{fig1_examples} 
\end{figure}

Salts in natural environments create spectacular landscapes but can also pose significant threats to ecosystems and historical artifacts\cite{pitman2002global,cooke1981salt,shahidzadeh2010}. Salts can dissolve in water and be transported over long distances, precipitating again where the water evaporates. This can occur, for example, when the water table rises by capillarity to the surface of soils ( granular materials ) to evaporate or when evaporation occurs near salt lakes. This is the origin of salt creeping, a phenomenon where salt crystals continue to precipitate far from an evaporating salt solution boundary\cite{washburn2002creeping,huang1976creeping,hird2016migration,qazi2019salt}. Qazi et al. demonstrated the self-amplifying mechanism of salt creeping far from the solution source. Due to multiple nucleation sites at the evaporation front , the spreading of the salt solution is enhanced well beyond the initial liquid/air front of the solution\cite{hird2016migration,qazi2019salt}. This, in turn, enlarges the evaporative area, leading to faster precipitation; this creates a self-amplifying process. The process results in three-dimensional crystalline networks at macroscopic distances from the salt solution source \cite{qazi2019}. Such crystallization process can initiate and grow on flat surfaces known as salt creeping and on the surface of  porous materials such as soil or stones, known as salt efflorescence \cite{desarnaud2015,qazi2019,veran2012discrete}. The process results in spectacular structures in natural environments, such as the formation of desert roses (assemblies of $CaSO_4$ or $BaSO_4$ crystals) in arid regions, salt pillars or other spectacular morphologies near the Black and Dead Sea coasts, and in Ethiopia's Danakil Depression (see Figure \ref{fig1_examples}a) \cite{shahid2009}. Additionally, salt creep can lead to soil salinization, vegetation decline, and water quality issues, impacting biodiversity \cite{pitman2002global}. Furthermore, the salt's ability to 'creep' causes damage in areas further than expected and can also damage outdoor electronics \cite{hienonen2000} and built cultural heritage, leading to costly restoration efforts. 
The efflorescence of the most ubiquitous salt on Earth, Sodium Chloride (NaCl), is commonly referred in the literature as 'cauliflower-shaped morphology' or 'patchy' efflorescence due to its resemblance to cauliflower or broccoli (see Figure \ref{fig1_examples}b) \cite{eloukabi2013,veran2014,bergstad2016,shahidzadeh2008}. It has been reported that these cauliflower structures, once formed on top of  porous/granular materials, are also porous and can subsequently absorb liquid by capillarity from the partially saturated porous materials upon evaporation  (Figure \ref{fig1_examples}c) \cite{eloukabi2013,sghaier2009,lazhar2020}. 

Here, we report on the microscale structure of precipitated salt as efflorescence and show that salt creeping induces the formation of a hierarchical porous medium. The internal porous structure is made of crystals with a fractal structure characterized by a self-organized, self-similar arrangement of salt crystals. Scanning Electron Microscopy (SEM) images of (NaCl) efflorescences reveal a structure similar to cauliflowers and show that this phenomenon covers many orders of magnitude in length scales, appearing fractal \cite{wijnhorst2024}. Our results show that the first precipitating large crystals create large pores between them, enabling smaller crystals to nucleate within these spaces, forming smaller pores, and so on. The hierarchicalporous structure of the precipitated salt cluster, with different pore size distributions, allows salt water to be absorbed by capillary pumping, enabling crystallization to continue even further; i.e. the salt solution continue to creep on its own precipitated crystals. The regularity in the crystal size has enabled us to create a predictive model, showing that the competition between capillarity and viscous effects does not limit the efflorescence height. The only limiting factor is the salt source, that is the amount of available salt solution. Our findings reveal how structures such as desert roses or salt pillars can grow to such macroscopic heights with a self-organized strucutre in nature.

For this purpose, we grown NaCl efflorescence (a) from almost saturated salt solution (5.5 M) in a hydrophobic (borosilicate) glass beaker to mimic the case of evaporation from salt lakes. Hydrophobic treatment based on silanisation is needed to suppress salt creep on hydrophilic glass walls of the beaker and induce preferential salt creeping on macroscopic precipitated salt crystals during evaporation (Figure \ref{fig1_examples}d). In addition, (b) we grow efflorescence on top of sandstone for situations where salt grows on top of porous materials saturated with salt solution (Figure\ref{fig1_examples}d). For the latter, before and after liquid immersion, the sandstone samples were weighed followed by sealing with three layers of parafilm on the lateral sides and bottom base to ensure drying solely from the top surface. Cone-shaped sandstone samples were used to facilitate the growth of larger efflorescence structures at an earlier stage of evaporation due to the high number of Péclets that is a characteristic of evaporation in truncated shapes \cite{wijnhorst2024}.  The volume of the truncated cone is $9.2 cm^{3}$, with a porosity ($\epsilon$) of 30\% and an average pore diameter ($d$) of 30 $\mu m$. The height of the cones is 50 mm, with a small base with radius $r_{s}$ of 5 mm at the top and a larger base with radius $r_{l}$ of 10 mm at the bottom. For the evaporation experiments from porous sandstone, we used an undersaturated NaCl solution of 1.7M, where $s = m_{i}/m_{s} = 0.3$ ($m_{i}$ denoting salt concentration and $m_{s}$ indicating saturation concentration) (Sigma Aldrich 99\%) \cite{wijnhorst2024}. The samples were saturated with the salt solution under vacuum for 1 hour. 

Both types of experiments (a and b) were dried in a controlled environmental chamber with an RH of 45$\pm$5$\%$ and a room temperature of 21$^\circ$C. It should be noted that relative humidity is an important factor in the formation of efflorescence as cauliflower. A low relative humidity (RH 20\%) modifies the morphology of the efflorescence by inducing horizontal growth of thin crystalline films as a crust \cite{desarnaud2015, qazi2019}  while a very high relative humidity favors the equilibrium growth morphology towards cubic crystals as cauliflower clusters.

\begin{figure}
\centering
\includegraphics[width=1\textwidth]{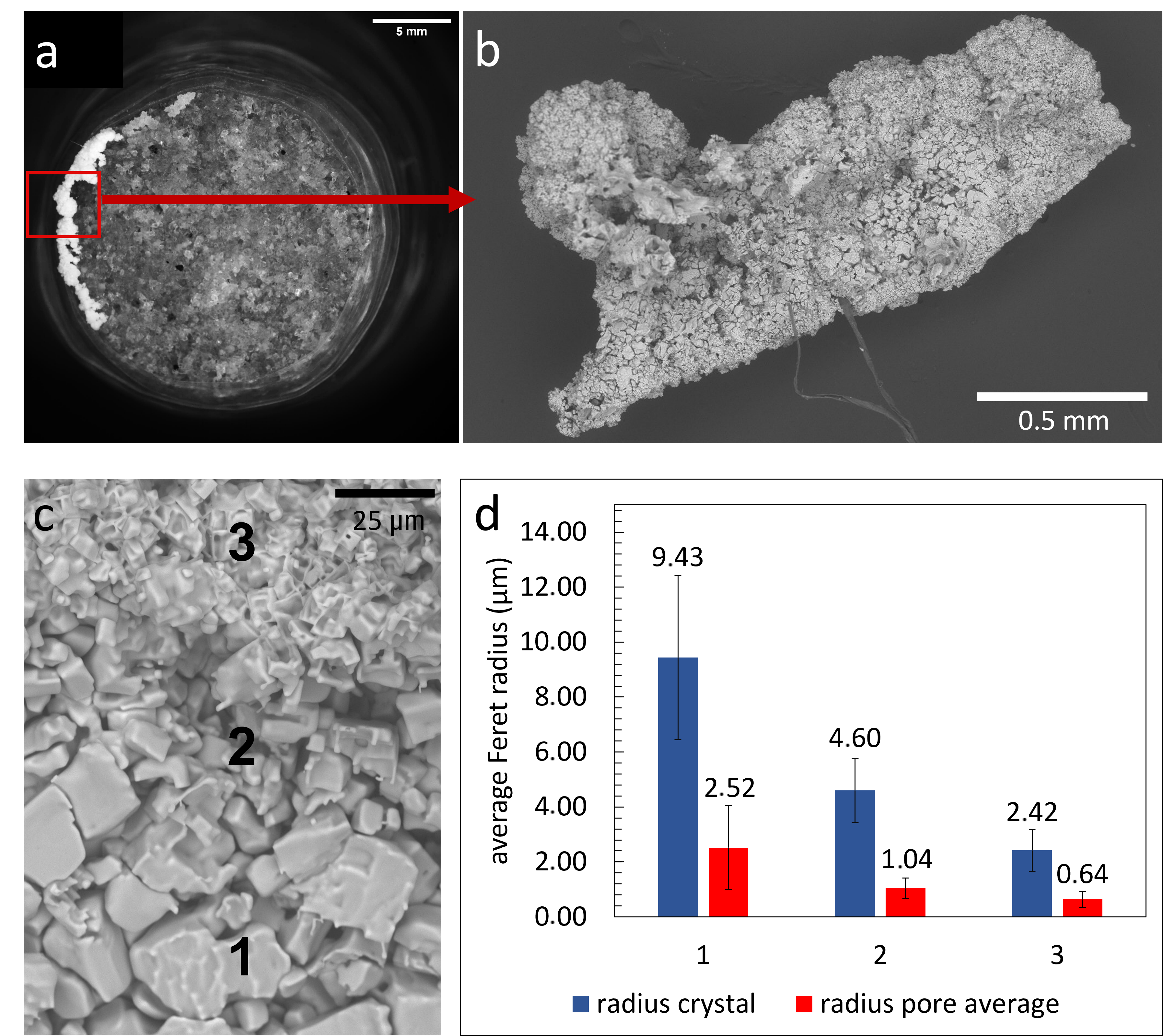}
  \caption{a) top view of the surface of the sandstone , early stage of efflorescence deposit of figure 1c.b) SEM image of a piece of the NaCl efflorescence.c) Hierarchical structure within the efflorescence with three layers of microcrystals with different sizes. d) shows the box-plot depicting the average Feret diameter of the pores per layer and the diameter of the pores per layer for 5 experiments.}
  \label{fig2_structure}
\end{figure}

\begin{figure}
\centering
\includegraphics[width=1\textwidth]{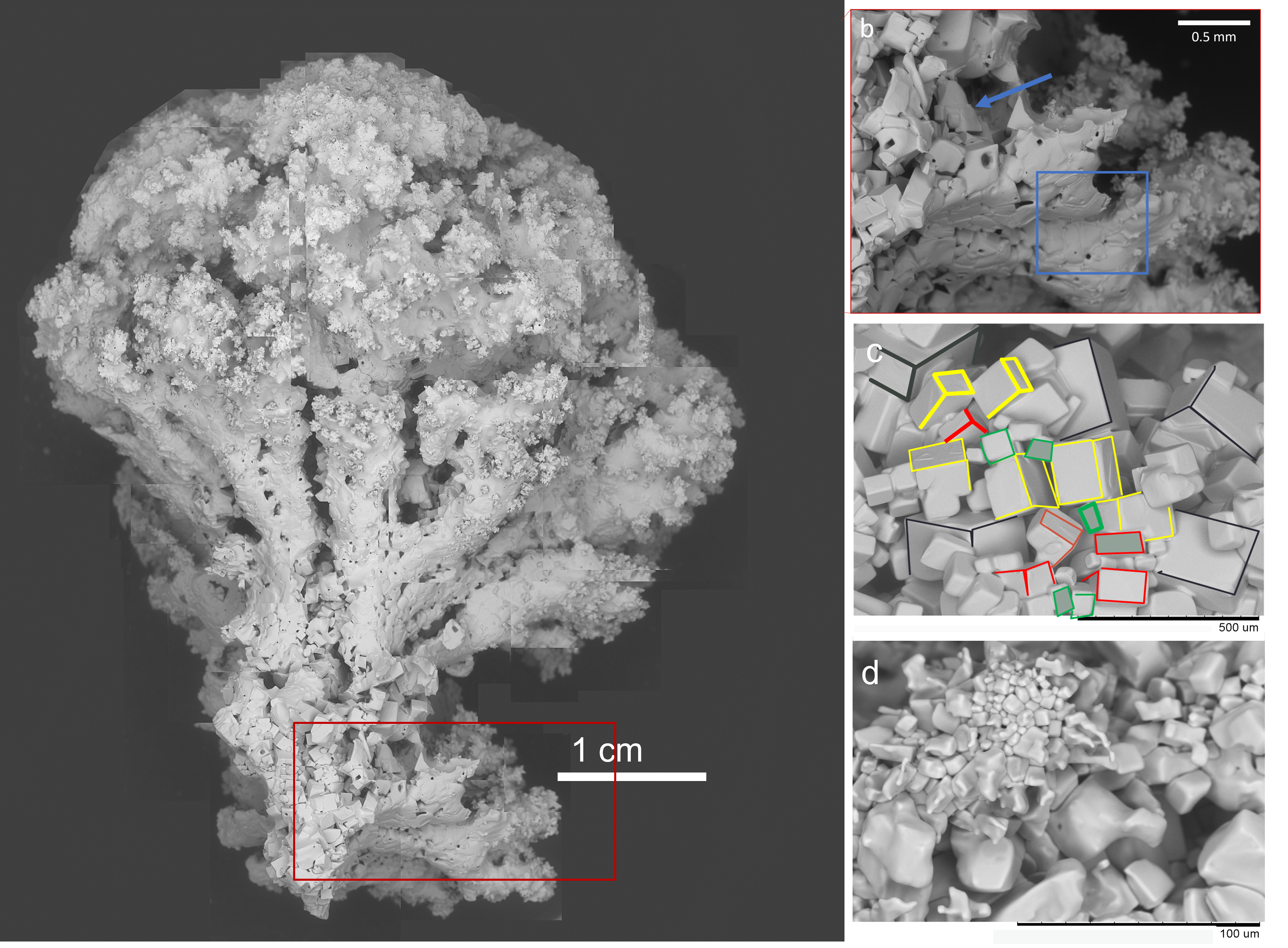}
  \caption{a) Scanning electron microscopy image of an emergent NaCl creep structure (pillar of salt) that looks like a cauliflower. b) Zoom of the emergent NaCl creep structure: cubical building blocks on the inner part (blue arrow) and skin formation on the outer part (blue box). c) the structure of the inner part of the base of the cauliflower showing the size dependent precipitation in the porous structure of the cauliflower shape deposit d) top view at the top surface : again distinct layers with crystals with different sizes are discernible }
  \label{fig3_cauliflower}
\end{figure}

With the evaporation of the salt solution, saturation concentration is reached followed by crystals precipitation as growing efflorescence . On top of porous materials, these localized structures form at the early stage of drying. In bulk salt solution experiments, the growth of efflorescence as macroscopic salt pillars initiates at a later stage of evaporation when the level of the salt solution reaches the upper surface of the precipitated NaCl cubed-shaped macro crystals. (Figure \ref{fig1_examples}d). The thin films of salt solution wetting the macroscopic NaCl crystals initiates the precipitation of smaller crystals by salt creeping mechanism leading to cauliflower-shaped efflorescence, composed of an assembly cubical micro crystals (Figure \ref{fig3_cauliflower}d) , i.e.like a house of cards. The efflorescence structures and the salt pillars were gently removed for examination using scanning electron microscopy (TM300 tabletop microscope from Hitachi (SEM)) (Figure \ref{fig2_structure}and\ref{fig3_cauliflower}) .

 Figure \ref{fig2_structure}c reveals that the hierarchically ordered micro crystals that collectively form the emergent cauliflower shape span very different length scales. The internal microscale morphology is composed of an assembly of cubic microcrystals (blue arrow in Figure \ref{fig3_cauliflower}b). When analyzing the efflorescence from the bottom to the top, different distinct layers with crystals of different sizes are discernible. The first precipitating large crystals create large pores between them (indicated in black in Figure \ref{fig3_cauliflower}c), enabling smaller crystals (yellow cubes) to precipitate within this space in the next layer. These crystals in turn form smaller pores which will induce the precipitation of even smaller crystals  (red and green in the figure), creating the next outer surface layer of salt, and so on (Figure \ref{fig3_cauliflower}). It is interesting to note that the final extreme outer surface ends with very thin salt skin with very small pores (blue square in Figure \ref{fig3_cauliflower}b).

\begin{figure}
\centering
\includegraphics[width=1\textwidth]{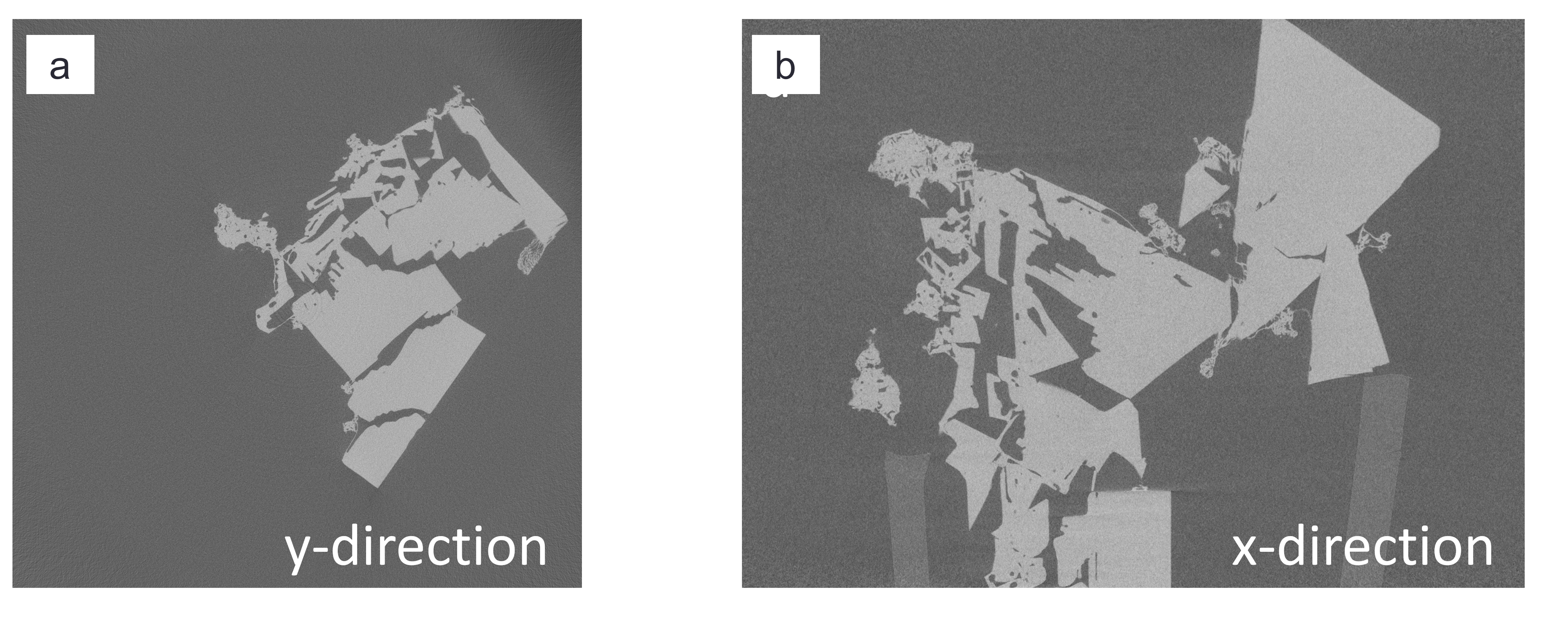}
  \caption{Micro-CT image of a cauliflower shaped NaCl crystal formed by creep showing the hollow parts inside the NaCl structure a) shows a slice taken in the y-direction, b) shows a slice taken in the x-direction}
  \label{fig4_Tomography}
\end{figure}

By measuring the Feret diameter of the salt cubes and the diameter of the pores in each layer using ImageJ software, the hierarchical relationship between the layers becomes evident. The boxplot in Figure \ref{fig2_structure}d demonstrates that in each layer, the average cube size is approximately half that of the previous layer, with a similar relationship observed for pore sizes. The pore sizes are roughly a quarter of the cube sizes in each layer, corresponding to a porosity ($\epsilon$) of approximately 20\%. 

Furthermore, we binarized the scanning electron microscopy images to calculate the fractal dimension of the efflorescence structure in 2D using the box-counting method \cite{mandelbrot1982}. Box counting is a technique used to estimate the fractal dimension of an object. In our case, a grid of equally sized squares is placed over the binary image of the efflorescence structure. The grid size is defined by a parameter $\varepsilon$, which is the length of the side of each square. Subsequently, the number of boxes that contain a part of the fractal is counted $N(\varepsilon)$. This is repeated for different values of $\varepsilon$  for the squares. The fractal dimension $D$ is estimated by analyzing how the number of squares needed to cover the pattern ($N(\varepsilon)$) changes with the length scale of the square ($\varepsilon$). 
If the structure is self-similar, the relationship between $N(\varepsilon)$ and $\varepsilon$ follows a power law:
 \[ N(\varepsilon) \propto 	\varepsilon^{-D} \] 
Taking the logarithm of both sides and plotting the resulting $\log(N(\varepsilon))$ against $\log(\varepsilon)$ should give a straight line. The slope of this line is the fractal dimension $D$.  
Using the box counting method function in ImageJ to perform this analysis we got an average fractal dimension of $1.71 \pm 0.03$ for the binarized 2D images of the NaCl structures. A fractal dimension of 1.7 suggests a relatively complex structure with a high degree of irregularity or roughness. The closer the fractal dimension is to 2, the more complex and space-filling the object becomes.  A dimension of 1.7 is quite common in nature ; coastlines, leaf structures, or certain geological formations, often have fractal dimensions between 1 and 2.

The results obtained by SEM imaging were completed by Micro-CT analysis. The latter provided more information on the internal 3D structure of the cauliflower efflorescence. The hierarchical crystalline structure is again visible (Figures \ref{fig4_Tomography}a and b). Some of the cubic crystals appear to be hollow on the inside. The 3D Micro-CT scans enables us to calculate the fractal dimension of NaCl creep in 3D by using a z-stack of the Micro-CT slices and applying a similar box-counting method, now programmed in Python\cite{mandelbrot1982}. We obtained a fractal dimension of approximately 2.562 ± 0.0013. This dimension corresponds to other self-similar branch-shaped structures in nature, such as real cauliflowers (2.8) and broccoli (2.7) \cite{kim2004}. Furthermore, the fractal dimensions obtained from the 2D images (1.71) and the 3D scan (2.5) are similar to those obtained for diffusion-limited aggregation (DLA) processes \cite{witten1983,halsey2000}.  The latter consists of aggregates formation in a system where particles are randomly moving by Brownian motion and stick to the growing aggregate when they come into contact with it. The resulting structure also forms a fractal-like pattern, often resembling branching or dendritic structures. The reason why NaCl efflorescence forms DLA-like structures is an interesting question for further study. 

\begin{figure}
\centering
\includegraphics[width=1\textwidth]{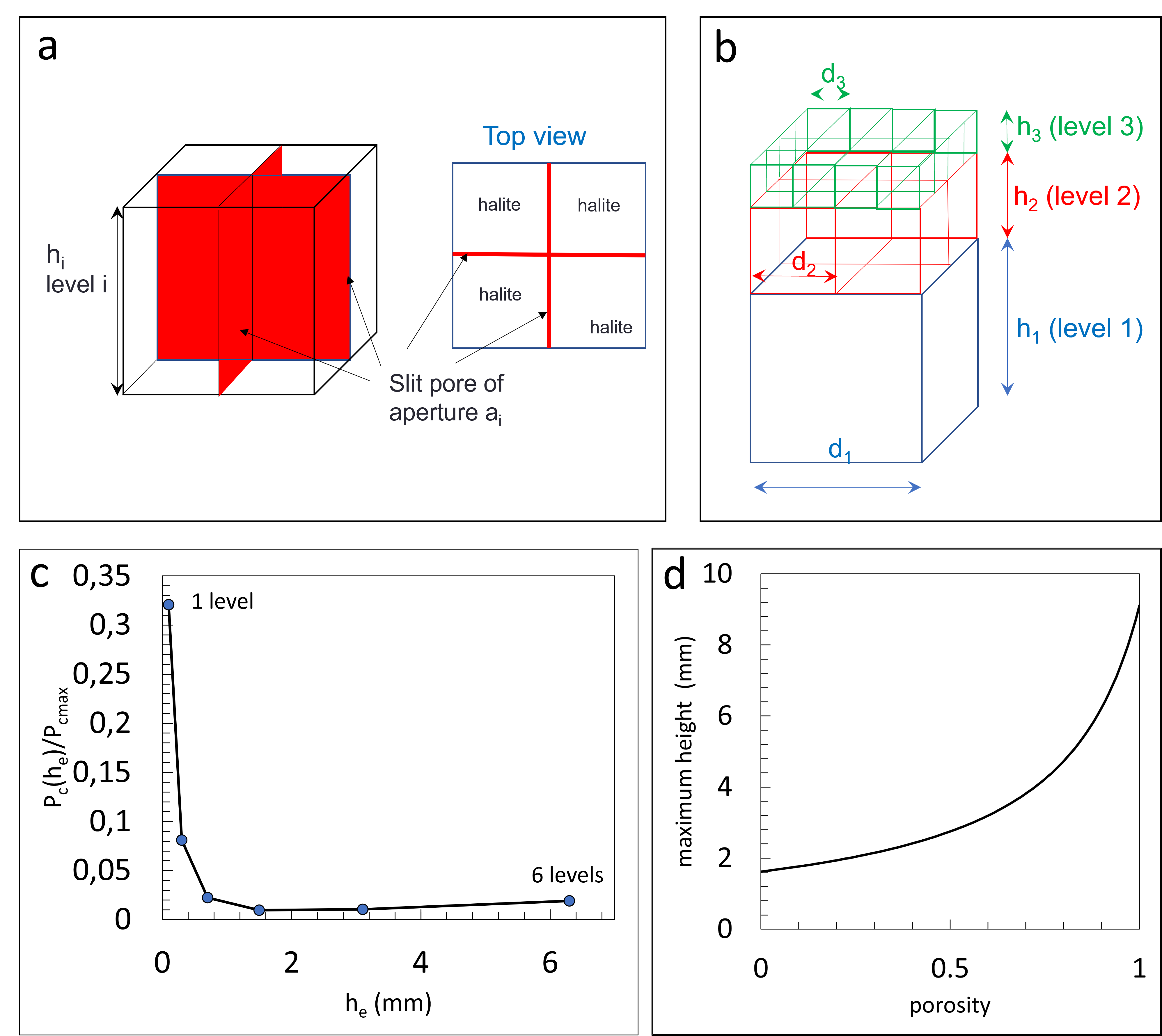}
\caption{a) Model of one level $h_{i}$ of NaCl (halite) with slit pores of aperture $a_{i}$. b) Hierarchical model with three levels of cubes with diameters $d_{i}$ in each layer n for 3 layers. c) Ratio of capillary pressure at maximum height $h_{e}$ to the maximum capillary pressure for n levels (for n=1 to n=6) and the corresponding maximum height. d) maximum height as a function of efflorescence porosity when the efflorescence height is limited by the amount of salt introduced in the experiment.}
\label{fig5_model}
\end{figure} 

From the porous structure of the salt efflorescence, one would expect that the competition between capillary pressure and viscous pressure drop should impact the maximum height of the efflorescence structure\cite{eloukabi2013,sghaier2009,lazhar2020}. If the capillary pressure below the efflorescence or creep structure (for example, inside a porous stone) ($P_{cpm}$) is smaller than the capillary pressure inside the efflorescence itself ($P_{ce}$), the porous efflorescence becomes saturated with salt solution due to capillary rise. On the other hand, evaporation at the surface of the efflorescence induces a viscous flow that creates a viscous pressure drop $\Delta P$ over the total height of the porous efflorescence. If $P_{ce}-P_{cpm} < \Delta P$, the capillary suction would be insufficient to maintain efflorescence growth, and the maximum height of the efflorescence would be reached. Increasing the evaporation rate would increase the pressure drop and result in lower efflorescence \cite{du1996,eloukabi2013,gupta2014}.

Our micro-scale investigation allows us to consider the NaCl efflorescence as a hierarchical porous medium (see Figure \ref{fig5_model}a), where each layer i consists of an assembly of cubical blocks with sides of $d_{i}$. Slits with an aperture of $a_{i}$ represent the pores in each layer. The height of each layer is denoted as $h_{i}$. Figure \ref{fig5_model}b depicts the situation where $h_{i} = d_{i}$, but $h_{i}$ can be larger than $d_{i}$, resulting in several cube sizes over the thickness of each layer. The total length of each layer is always equal to $d_{1}$. As our observations show that for each layer the cube sizes are devided by two as the pore sizes we get the following relationships:
\begin{enumerate}
    \item $d_{i} = d_{i-1}/2 $
    \item $a_{i} = a_{i-1}/2 $
    \item $h_{i} = \alpha h_{i-1}$
\end{enumerate}

From Darcy's law, we derive the viscous pressure drop over level $i$ ($p_{i}$):
$\Delta p_{i}=\frac{\mu jh_{i}}{\rho_{l} k_{i}}$
where $\mu$ is the viscosity of the salt solution, $j$ is the evaporation flux, and $\rho_{l}$ is the density of the NaCl solution.

The viscous pressure drop ($\Delta P$) for flow perpendicular over n parallel levels is thus  given by the sum of  pressure drops in each level: 
\begin{equation*}
\Delta P = \sum_{i=1}^{i=n} \Delta p_{i} =\sum_{i=1}^{i=n}\frac{\mu jh_{i}}{\rho _{l} k_{i}}
\label{deltap}
\end{equation*} 
The expression for the permeability per level $k_{i}$ is given by the so called cubic law that considers flow through rock fractures as considering the fracture as an aperture $a_{i}$ between two parallel plates with length $d_{1}$ \cite{zimmerman1996}: $k_{i}=2^{i} \frac{a_{i}^{3}}{12d_{1}}$
By filling in this expression for the permeability ($k_{i}$) of each layer $i$ in equation \ref{deltap} we get the following expression for the pressure drop:
\begin{equation*}
\Delta P =\frac{12d_{1}\mu j}{\rho_{l}} \sum_{i=1}^{i=n}\frac{2^{3i-3} h_{i}}{2^{i} a_{1}^{3}} = \frac{12d_{1} \mu j}{\rho_{l} a_{1}^{3}} \sum_{i=1}^{i=n}2^{2i-3} h_{i}
\end{equation*} 

The maximum capillary pressure $P_{ce}$on top of the n levels is given by the Young-Laplace equation\cite{gennes2004}:
\begin{equation}
P_{ce}(n)=\frac{2cos\theta \gamma}{a_{n}} =\frac{2^{n}cos\theta \gamma}{a_{1}}
\end{equation}
where $\theta$ is the contact angle between the NaCl solution and the NaCl crystals forming the efflorescence (considered as 0 in this case), and $\gamma$  is the surface tension of the NaCl solution ($\sim 75 mN/m$ \cite{wang2018}). 

Now the last pressure that needs to be defined is the pressure inside the porous medium underneath the salt crust $P_{cpm}$. For that we use the retention curve for drying and growing NaCl as obtained by 
Hidri et al\cite{hidri2013}. A retention curve shows the relationship between the amount of liquid a porous medium can hold for  a certain pressure: 

\begin{equation}
\frac{P_{cpm}}{P_{cref}} = {\left({\left(\frac{(1-S_{C})}{(S-S_{C})}\right)}^{\frac{1}{m}}-1\right)}^{\frac{1}{n}}
\label{vangen}
\end{equation}

With $S$ the liquid saturation, $S_{c}$ the residual liquid saturation $n$ a measure of the pore size distribution and $m = 1 - 1/n$. $P_{cref}$ is the capillary entry pressure, and is classically referred to as the minimum pressure difference between the two fluids that allows the breakthrough of the non-wetting phase through a porous sample. The capillary entry pressure plays a key role in controlling saturation of the porous medium. $P_{cref}$ is defined as: $\frac{6(1 - \epsilon)\gamma}{\epsilon d}$\cite{hidri2013}. Here, $\epsilon$  is the porosity of the porous stone (0.28), and d is the porous medium equivalent grain size (100 $\mu$m). 
Hidri et al.\cite{hidri2013} got the value of $S_{c}=0.1$ from an experimental study performed by Dullien et al. \cite{27}. They achieved $n=10$ by fitting the parameters of the classical relationship described in equation \ref{vangen} with the data from Dodds et al. \cite{28}.

Using these formulations for the viscous pressure drop and the capillary pressure in the porous stone, we can derive the capillary pressure on top of the efflorescence (at $h_{e}$) in our experiments, where the efflorescence consists of n layers with different cube sizes:

\begin{equation}
P_{ce}(h_{e})=\frac{12d_{1}\mu jh_{1}}{8\rho_{l}a_{1}^{3}}\left(\frac{4^{n+1}\alpha^{n}-4}{4\alpha-1}\right)+P_{cpm}
\end{equation}

Where $h_{e} = h_{1}(\alpha^{n} - 1)/(\alpha - 1)$, and the height of the first level is taken as $h_{1} = 0.1 mm$, which is about 10 level 1 cube sizes in our experiments. The computation is performed for $h_{i} = 2h_{i-1}$, meaning that the thickness of level i is twice the thickness of level $i-1$ (i.e., $\alpha$ = 2). The evaporation flux in our experiments was $j = 8.5 . 10^{-5} kg/m^{2}/s$. In Figure \ref{fig5_model}c, $P_{c}(h_{e})/P_{cmax}$ is plotted as a function of the maximum height for cases from n = 1 to n = 6. As we can see from this graph, $P_{c}(h_{e})/P_{cmax} < 1$, indicating that the efflorescence growth cannot be limited by capillary-viscous competition. To investigate the hypothesis of initial salt mass as the limiting factor, we equate the mass of salt in the solution ($m_{s}$) at the beginning of the experiment in the porous material test:
\begin{equation}
m_{s}=\rho_{l} C_{0}\epsilon V
\end{equation}
(where $C_{0}$ is the initial mass fraction of NaCl in the aqueous solution of 10\% and V is the volume of the sandstone sample and $\epsilon$ its porosity) with the mass of salt present in efflorescence with height $h_{e}$:
\begin{equation}
m_{s} = A h_{e} (\rho_{cr}(1-\epsilon_{e})+\rho_{l} C_{sat} \epsilon_{e})
\end{equation}
(where $A$ is the evaporation surface area of the sandstone sample, $\rho_{cr}$ is the density of NaCl crystals ($2.17 g/cm^{3}$) \cite{smith1976}, $\epsilon_{e}$ is the porosity of the efflorescence, and $C_{sat}$ is the solubility mass fraction (0.36 for NaCl)). We derive the following formula for the maximum height of the efflorescence ($h_{e}$) as a function of its porosity ($\epsilon_{e}$):
\begin{equation}
h_{e}=\frac{\rho_{l} C_{0}\epsilon}{(\rho_{cr}(1-\epsilon_{e})+\rho_{l} C_{sat} \epsilon_{e})} \frac{V}{A}
\end{equation}
Figure \ref{fig5_model}d shows that the resulting maximum heights align with the order of magnitude of the maximum heights observed in our experimental results.

In summary,  we show that NaCl efflorescence structures formed during evaporation exhibit a fractal geometry similar to cauliflower and broccoli. The precipitated salt cluster exhibits self-similarity: identical patterns can be observed at increasingly smaller scales; the crystal growth process repeats itself at multiple scales, with smaller versions of the same structure appearing on the larger ones.   Microscale investigation shows that the cauliflower-like structure of efflorescence is built up of an assembly of micro-crystals that together can form complex structures with much larger surface area relative to its volume; The hierarchical build-up of the crystals forms a fractal structure with a fractal dimension of 1.71 for the 2D images and 2.5 for the 3D micro-CT scan. The fractal pattern emerges from the repeated application of similar crystal growth rules, where each new growth point follows the same pattern as the whole with similar levels of complexity and space-filling properties. We developed a hierarchical model for the efflorescence structure to explore the capillary-viscous efflorescence height limit. The model indicates that the capillary-viscous competition is not the limiting factor. In our experiment, the efflorescence height is only limited by the amount of salt solution used at the beginning of the experiment. However, in some conditions the structure could partially collapse under its own weight when the efflorescence is wet and the capillary adhesion between the microcrystals is weak. 
Our findings reveal how mineral deopsits structures such as desert roses or salt pillars can grow to macroscopic heights in nature due to their internal fractal geometry, leading to a porous structure with very small pores. 
Other examples of fractal-like structures in nature are coastlines, river systems, mountain ranges, and biological systems like the branching of trees or blood vessels . In addition, various parts of fruits, from their surface textures to their internal and external patterns can reveal self-similar behavior.

Further investigation is needed to explain these growth structures on an ionic scale, to be able to further model and predict the development of the 'cauliflower NaCl efflorescence' in three dimensions for a better understanding of crystal growth in porous media and in nature.

\begin{acknowledgement}
The authors would like to thank N.Schramma and P.Kolpakov for their help with some experiments and the analysis. 

\end{acknowledgement}








\bibliography{achemso-demo}

\end{document}